\documentclass[reprint, showpacs, superscriptaddress]{revtex4-1}
\usepackage[utf8]{inputenc}
\usepackage[english]{babel}
\usepackage{amsmath}
\usepackage{amsfonts}
\usepackage{amssymb}
\usepackage{graphicx}
\usepackage{dcolumn}
\usepackage{booktabs}
\AtBeginDocument{
\heavyrulewidth=.08em
\lightrulewidth=.05em
\cmidrulewidth=.03em
\belowrulesep=.65ex
\belowbottomsep=0pt
\aboverulesep=.4ex
\abovetopsep=0pt
\cmidrulesep=\doublerulesep
\cmidrulekern=.5em
\defaultaddspace=.5em
} 
\begin{document}
\author{B.G.C. Lackenby}
\affiliation{School of Physics, University of New South Wales,  Sydney 2052,  Australia}
\title{Calculation of atomic spectra and transition amplitudes for superheavy element Db (Z=105)}
\author{V.A. Dzuba}
\affiliation{School of Physics, University of New South Wales,  Sydney 2052,  Australia}
\author{V.V. Flambaum}
\affiliation{School of Physics, University of New South Wales,  Sydney 2052,  Australia}
\affiliation{Johannes Gutenberg-Universit\"at Mainz, 55099 Mainz, Germany}
\begin{abstract}
Atomic spectra and other properties of superheavy element dubnium (Db, Z=105) are calculated using
recently developed method combining configuration interaction with perturbation theory (the CIPT method,
Dzuba et al, Phys. Rev. A, {\bf 95}, 012503 (2017)). These include energy levels for low-lying states of Db and Db~II,
electric dipole transition amplitudes from the ground state of Db, isotope shift for these transitions and ionisation 
potential of Db. Similar calculations for Ta, which is lighter analog of Db, are performed to control the accuracy
of the calculations.
\end{abstract}

\maketitle

\section{Introduction}

The study of super heavy elements (SHE), $Z > 100$, which are not found in nature has been a constant frontier of nuclear and atomic physics for the past century \cite{Oganessian2009,  HHO2013}. The discovery and study of these exotic elements is of great interest at both microscopic and cosmological scales, particularly in relation to the existence of the fabled island of stability which is the shell model's promise of stable super heavy elements and the property of the underlying exotic nuclei \cite{OUL2004, HHO2013, Leino2016, Oganessian2012}.  The study of SHEs also gives insight into fundamental physics such as correlation and relativistic interactions in atomic systems. \\

Despite the rich potential for studying new physics, both the experimental and theoretical knowledge of SHEs is poor. 
While elements up to $Z$=118 have been synthesized, studying properties of these elements, particularly their spectra, 
is exceptionally difficult due to short lifetimes and low production rate.  Currently experiments are underway to study 
these elements in greater detail, including measuring their atomic spectra. So far there has been success in measuring 
the $^1$S$_0 \rightarrow ^1$P$^{\rm o}_1$ excitation energy of No  ($Z=102$) \cite{Laatiaoui2016} and ionisation potentials
of No~\cite{Laatiaoui2016} and Lr ($Z=103$) \cite{SAB15}. The development and refinement of laser spectroscopy 
techniques make future experimental measurements in the SHE region promising 
\cite{Laatiaoui2014, Laatiaoui20161, Ferrer2017}.  The theoretical results presented here will facilitate future experiments. \\

There has been significant study of SHEs with a small number of electrons (holes) above (below) closed shells. These calculations have been performed using well-established many-body techniques such  as couple-cluster methods \cite{Lindgren1986, Blundell1991}, CI + MBPT \cite{Dzuba1996}, correlation potential (CP) methods \cite{Dzuba1989} and Multiconfigurational Dirac-Fock (MCDF) \cite{Grant1988} etc.  For SHEs $Z=102,103,104$ which have 2, 3, and 4 valence electrons above the  closed $f$ shell, their spectra, ionisation potentials and static polarisabilities have been calculated \cite{Liu2007, Desclaux1980, Eliav1995, Fritzsche2007, Zou2002, Borschevsky2007, Martin1996, Mosyagin2010, DSS2014, Eliav2015, Kaldor2007}. Similarly atomic properties of SHEs  $Z=112-118$ using  coupled-cluster methods\cite{DF2016, Pershina2008, Nash2005, Landau2001, Borschevsky2015, Thierfelder2008, Kaldor2008, Dinh2016}, CI+MBPT methods \cite{Dinh2008, Dzuba1996} and MCDF methods\cite{Eliav1996, Yu2008} have been calculated. Atomic properties of unsynthesized nuclei have also been theoretically calculated up to the SHE element $Z=122$~\cite{Dinh2008a, Ginges2015, Eliav2015}. A review of SHE atomic calculations can be found  in ref. \cite{Eliav2015}.\\
 
While established numerical methods have been successful in treating SHE with relatively simple atomic structure, difficulties arise when there are more than four valences electrons in the open subshells in many-electron atoms ($Z=105-111$) due to the extremely large CI basis and therefore calculations have been limited to ionisation potentials and electric polarizabilites \cite{Turler2013,  Dzuba2016, Johnson1999, Johnson2002}.  A Recently developed method combining configuration interaction with perturbation theory (the CIPT method~\cite{DBHF2017}) allows to overcome these limitations and perform the calculations 
for the rest of SHE. In this work we start this study with the calculation of  atomic spectra of Ta I and Db I which both have five valence electrons above a closed $f$ shell. 
Calculations for neutral Ta atom ($Z=73$) are performed to demonstrate the accuracy and practicality of the calculations.  \\

In Section \ref{sec:CIPT} we briefly discuss the application of the CIPT method, in Section \ref{sec:TaI} we compare our CIPT calculations for Ta I with experimental results. We present the CIPT calculations for Db I in Section \ref{sec:DbI} including calculations of Breit and radiative corrections. Finally in Section \ref{sec:E1transitions} we present the optical electric dipole transitions for both Ta I and Db I including calculations of the isotope shift for Db I.  \\

\section{The CIPT Method} \label{sec:CIPT}

The sparse theoretical results for elements $Z=105-111$ is due to the open $6d-$shell where current models are not viable. For more than four valence electrons previous many-body methods become too computationally expensive due to the large diagonalisation problem. The computational cost is reduced by using a combination of the configuration interaction (CI) and perturbation theory (PT). In this section we will give a brief discussion of our implementation of the CIPT method. For an in depth discussion please refer to Ref. \cite{DBHF2017}. 

To generate the single electron wavefunctions for both Ta  and Db  we use the $V^{N-1}$ approximation
($N$ is the total number of electrons) \cite{Kelly1964, Dzuba2005}, 
starting the Hartree-Fock calculations for an open-shell atom with the $5d^36s$ configuration of external electrons 
for Ta and the $6d^37s$ configuration For Db.  Single-electron basis states are calculated in the field of frozen core.
We a B-spline technique~\cite{Johnson1988}  with 40 B-spline states of order 9 in a box with radius $40 \ a_B$ 
with partial waves up to $l_{max}=4$. Many-electron basis states for the CI calculations are formed by making
all possible single and double excitations from reference low-lying configurations. Note that this choice of the
$V^{N-1}$ potential is most natural for Ta where most of excited configurations contain exitations of the $6s$
electron. However, it is not so for Db. Here due to strong relativistic effects the $7s$ electrons are attracted to the
core and the $6d$ electrons are excited instead. Therefore, for Db for also tried a $V^{N-1}$ potential calculated
for the $6d^27s^2$ configuration of external electrons. The difference in results turned to be small. \\

In this work we include the effects of both the Breit interaction \cite{Breit1929, Mann1971} and radiative corrections (self-energy and vacuum polarisation corrections) \cite{FG2005} for completeness. It is expected that these corrections are more significant in heavier elements than their lighter element analogs. The Breit interaction which accounts for the magnetic interaction and retardation is included in the zero momentum transfer approximation,
\begin{align*}
\hat{H}^B = -\dfrac{\boldsymbol{\alpha}_1 \cdot \boldsymbol{\alpha}_2 + \left(\boldsymbol{\alpha}_1\cdot\textbf{n}\right)\left(\boldsymbol{\alpha}_2\cdot \textbf{n}\right)}{2r}
\end{align*}
where $\boldsymbol{\alpha}$ is the Dirac matrix, $\textbf{r}=r\textbf{n}$ and $r$ is the distance between electrons denoted by subscripts $1$ and $2$. The QED radiative corrections due to the Uehling potential $V_U$, and electric and magnetic form-factors $V_E $ and $ V_g$ respectively are included. These corrections are included are effective radiative potentials as introduced in ref. \cite{FG2005},
\begin{align*}
V_{R}(r) = V_{U}(r) + V_{g}(r) + V_{e}(r).
\end{align*}
Both the Breit and effective radiation correction potentials are included  in the Hartree-Fock procedure described above. In Section \ref{sec:DbI} we present the individual effect of each correction to the SHE Db I spectrum from the CIPT method. \\

For each level we calculate the Land\'{e} g-factor and compare it to the non-relativistic expression,
\begin{align} \label{eq:Lande}
g =  1 + \dfrac{J(J + 1) - L(L+1) + S(S+1)}{2J(J+1)}.
\end{align}
We treat angular momentum $L$ and spin $S$ as fitting parameters to fit the calculated values of the $g$-factors
with the formula (\ref{eq:Lande}). This allows us to use the $LS$ notations for atomic states.
Note however that the SHE states are highly relativistic and strongly mixed and the $LS-$ coupling scheme 
is very approximate. 

Ionisation potential is obtained by calculating the energy of the ground state of the ion and taking the difference 
between ground states of the ion and the neutral atom. The same single-electron basis is used for the ion as for the
neutral atom. 

\section{Ta I } \label{sec:TaI}
\begin{table*}[p!]
\center
\caption{Comparison of experimental \cite{NIST_ASD} and CIPT spectra of Ta I. Here we have presented the first 5 even parity states and the first 40 odd parity states of Ta I. We have also included the a comparison of the ionisation energy and first two ionic states of Ta I.  The presented state is derived from the approximate g-factors generated by the CIPT code.  The final column is the difference between experimental and theoretical results $\Delta = E_{E} - E_{T}$. \label{tab:TaComparison}}
\begin{tabular}{cl@{\hspace{0.5cm}}l@{\hspace{0.5cm}}c@{\hspace{0.5cm}}r@{\hspace{0.5cm}}r@{\hspace{0.5cm}}r@{\hspace{0.5cm}}r@{\hspace{0.5cm}}r}
\toprule
\toprule
& & & \multicolumn{2}{c}{Experimental} & \multicolumn{2}{c}{CIPT} &  \\
\cmidrule{4-5} \cmidrule{6-7}
& Configuration & State & $J$ &  \multicolumn{1}{c}{$E_E$} &  \multicolumn{1}{c}{$g_E$} &  \multicolumn{1}{c}{$E_T$} &  \multicolumn{1}{c}{$g_T$} &  \multicolumn{1}{c}{$\Delta$} \\
\midrule
& Even states\\
(1)  &$5d^3 6s^2$ & $^4$F & 3/2 & 0.00 & 0.447 &  0.00 & 0.4373 & \\
(2)  &$5d^3 6s^2$ & $^4$F & 5/2 & 2 010 & 1.031 &  1 652 & 1.0336 & 358 \\
(3)  &$5d^3 6s^2$ & $^4$F & 7/2 & 3 964 & 1.218 & 3 175 & 1.2265 &  789\\
(4)  &$5d^3 6s^2$ & $^4$F & 9/2 & 5 621 & 1.272 & 4 679 & 1.3066 & 942\\
(5)  &$5d^3 6s^2$ & $^4$P & 1/2 & 6 049 & 2.454 & 6 017 & 2.4022 & 32\\
\\
& Odd states\\
(6)  &$5d^3 6s 6p$ & $^6$G$^{\rm_o}$  & 3/2 & 17 385 &  & 17 599 &   0.1719   &  -214\\
(7)  &$5d^3 6s 6p$ & $^2$F$^{\rm_o}$  & 5/2 & 17 994 & 0.732 &  18 225 &   0.7955  & -231\\
(8)  &$5d^2 6s^2 6p$ & $^4$D$^{\rm_o}$  & 1/2 & 18 505 & 0.172 & 18 629 &   0.0716  & -124\\
(9)  &$5d^3 6s 6p$ & $^6$G$^{\rm_o}$  & 5/2 & 19 178 & 0.851 & 19 393 &   0.8551 &  -123\\
(10)  &$5d^2 6s^2 6p$ & $^4$D$^{\rm_o}$  & 3/2 & 19 658 & 1.018 & 19 724 &   0.9389 & -66\\
(11)  &$5d^2 6s^2 6p$ & $^2$S$^{\rm_o}$  & 1/2 & 20 340 & 1.956 & 20 574 &   2.0278 & -233\\
(12)  &$5d^3 6s 6p$ & $^6$G$^{\rm_o}$  & 7/2 & 20 560 & 1.194 & 20 463 &   1.1394 & -97\\
(13)  &$5d^2 6s^2 6p$ & $^2$D$^{\rm_o}$  & 3/2 & 20 772 & 0.812 & 20 796 &     0.8124 & -24\\
(14)  &$5d^2 6s^2 6p$ & $^4$D$^{\rm_o}$  & 5/2 & 21 168 &   & 21 358 &   1.2117 &  -190\\
(15)  &$5d^3 6s 6p$ & $^4$F$^{\rm_o}$ & 3/2 & 21 855 & 0.666 & 22 132 & 0.6773 & -277 \\
(16)  &$5d^2 6s^2 6p$ & $^2$D$^{\rm_o}$ & 5/2 & 22 047 & 1.179 & 21 875 & 1.0838 & 172 \\
(17)  &$5d^2 6s^2 6p$ & $^4$G$^{\rm_o}$  & 7/2 & 22 381 & 1.060 & 22 276 & 1.0377 & 105 \\
(18)  &$5d^3 6s 6p$ & $^6$G$^{\rm_o}$ & 9/2 & 22 682 & 1.231 & 22 285 & 1.2677 & 397 \\
(19)  &$5d^3 6s 6p$ & $^6$F$^{\rm_o}$ & 1/2 & 23 355 & -0.320 & 23 680 & -0.2689 & -325 \\
(20)  &$5d^3 6s 6p$ & $^4$F$^{\rm_o}$ & 5/2 & 23 363 & 1.078 & 23 381 & 1.0766 & -18 \\
(21)  &$5d^2 6s^2 6p$ & $^4$D$^{\rm_o}$ & 7/2 & 23 927 & 1.326 & 23572 & 1.3256 & 355 \\
(22)  &$5d^3 6s 6p$ & $^6$F$^{\rm_o}$ & 3/2 & 24 243 & 1.126 & 24 463 & 1.1018 & -220 \\
(23)  &$5d^3 6s 6p$ & $^6$D$^{\rm_o}$ & 1/2 & 24 517 & 2.888 & 24 907 & 2.9261 & -390 \\
(24)  &$5d^3 6s 6p$ & $^6$D$^{\rm_o}$ & 3/2 & 24 739 & 1.620 & 25 143 & 1.6808& -404 \\
(25)  &$5d^3 6s 6p$ & $^4$F$^{\rm_o}$  & 7/2 & 24 982 & 1.235  & 24 922 & 1.2590 & 60 \\
(26)  &$5d^3 6s 6p$ & $^6$G$^{\rm_o}$ & 11/2 & 25 009 & 1.302 & 24 528 & 1.3366 & 481 \\
(27)  &$5d^3 6s 6p$ & $^6$F$^{\rm_o}$ & 5/2 & 25 181 & 1.239 & 25 267 & 1.2573 & -86 \\
(28)  &$5d^3 6s 6p$ & $^6$G$^{\rm_o}$ & 9/2 & 25 186 &  & 24 733  & 1.2540 & 453 \\
(29)  &$5d^3 6s 6p$ & $^4$D$^{\rm_o}$ & 1/2 & 25 513 & 0.028 & 25 697 & 0.0319  & -184 \\
(30)  &$5d^3 6s 6p$ & $^4$F$^{\rm_o}$  & 9/2 & 25 926 &  1.292 & 25 509 & 1.2970 & 417 \\
(31)  &$5d^2 6s^2 6p$ & $^4$P$^{\rm_o}$ & 5/2 & 26 220 & 1.338 & 26 298 & 1.2923 & -78 \\
(32)  &$5d^3 6s 6p$ & $^4$D$^{\rm_o}$ & 3/2 & 26 364 & 1.393 & 26 678 & 1.2676 & -314 \\
(33)  &$5d^3 6s 6p$ & $^6$F$^{\rm_o}$  & 7/2 & 26 586 & 1.356  & 26 299 & 1.315 & 287 \\
(34)  &$5d^2 6s^2 6p$ & $^4$P$^{\rm_o}$ & 3/2 & 26 590 & 1.576 & 26 759 & 1.6833 & -169\\
(35)  &$5d^3 6s^3 6p$ & $^6$D$^{\rm_o}$ & 5/2 & 26 795 & 1.416 & 26 815 & 1.4086 & -20 \\
(36)  &$5d^2 6s^2 6p$ & $^4$P$^{\rm_o}$ & 1/2 & 26 866 & 2.650 & 27 094 & 2.6189 & -228\\
(37)  &$5d^3 6s 6p$ &  $^4$F$^{\rm_o}$   & 7/2 & 26 960 & 1.223  & 26 787 & 1.2390 & 173 \\
(38)  &$5d^3 6s 6p$ & $^6$F$^{\rm_o}$  & 9/2 & 27 733 &  1.390 & 27 279 & 1.3590 & 454 \\
(39)  &$5d^3 6s 6p$ & $4$D$^{\rm_o}$   & 7/2 & 27 781 & 1.374  & 27 643 & 1.4658 & 138 \\
(40)  &$5d^3 6s 6p$ & $^6$G$^{\rm_o}$ & 11/2 & 27 783 & 1.351 &  27 376 & 3.3534  & 407\\
(41)  &$5d^3 6s^3 6p$ & $^4$D$^{\rm_o}$ & 5/2 & 28 134 & 1.394 & 28 337 & 1.3665 & -203 \\
(42)  &$5d^3 6s 6p$ & $^4$G$^{\rm_o}$   & 7/2 & 28 183 & 1.115  & 27 970 & 1.0421 & 213 \\
(43) &$5d^3 6s 6p$ & $^2$P$^{\rm_o}$   & 3/2 & 28 689 & 1.356  & 28 693 & 1.3052  & -4 \\
(44)  &$5d^3 6s 6p$ & $^6$D$^{\rm_o}$  & 9/2 & 28 767 &  1.337 & 28 414 & 1.4106 & 353 \\
(45)  &$5d^3 6s^3 6p$ & $^6$F$^{\rm_o}$ & 5/2 & 28 862 & 1.247 & 28 868 & 1.2678 & -6 \\
(46)  &$5d^3 6s^3 6p$ & $^6$D$^{\rm_o}$ & 1/2 & 29 902 & 2.994 & 30 323 & 2.9971 & -421 \\
& Ta I ionisation potential \\
(47)  &$5d^3 6s$& $5$F & 1 & 60 891 & 0.000 & 61 073 &    0.0235 &  -182\\
\bottomrule
\bottomrule
\end{tabular}
\end{table*}

To demonstrate the accuracy of the CIPT model we compare the theoretical and experimental spectra of Ta I. As Ta lies in the same group but one period higher, we believe theoretical accuracy for the Ta spectrum would indicate what accuracy we can expect for Db.
Electron states of neutral Ta have an open $5d$ shell, its ground state configuration is [Xe]$4f^{14}5d^3 6s^2$.
As the $6s$ electrons are easily excited, we should treat the atom as a system with five external electrons.
Note that a slightly more complicated atom, tungsten, which has one more external electron, was already successfully 
studied using the CIPT method~\cite{DBHF2017}. Therefore, we expect similar or better accuracy for Ta.
For low lying even parity states of Ta we used the basis states of the $5d^3 6s^2$, $5d^4 6s$ and $5d^5$  
configurations in the effective CI matrix. All other configurations, which were obtained by exciting one or two 
electrons from these configurations, were included perturbatively. 
Similarly for the odd parity states we used the states of the $5d^3 6s 6p$, $5d^2 6s^2 6p$ configurations
in the effective CI matrix, while other configurations are included perturbatively.

In Table \ref{tab:TaComparison} we present the comparison between experimental energies and $g$-factors and 
those calculated by the CIPT method.  We present a significant number of odd states to demonstrate the accuracy of the odd parity states particularly towards to end of the optical region. This is because the most promising experimental measurements are strong optical electric dipole (E1) transitions from the ground state to low lying excited states of different parity and it is important to include as many of these transitions as possible. To identify the correct states for comparison we use the experimental and theoretical Land\'{e} $g$-factors, when experimental $g$-factors were not available we used the next sequential state in the theoretical calculations. There was excellent agreement between the $g$-factors with the only significant difference for the odd state $J=1/2$ at $18 505$ cm$^{-1}$.  \\

There is good agreement between the experimental and theoretical energies particularly for the low-lying odd parity states which are important for measuring the electric dipole transitions (see Section \ref{sec:E1transitions}). For the odd parity states the largest discrepancy in energy was $\Delta = 453$ cm$^{-1}$ with most states having $|\Delta| \approx 100-400$ cm$^{-1}$. Our calculations also supported the existence of the $J=11/2$ level at $E_E=  27 783$ cm$^{-1}$ which is listed as ambiguous \cite{NIST_ASD}.  For the calculation of the Db I spectrum we expect a similar accuracy as seen in Ta I due to the analogous electronic structure.\\

\section{Db I} \label{sec:DbI}
Dubnium was first synthesized in 1968 with the longest living isotope of $^{268}$Db~\cite{Schadel2012, Oganessian2005} with a halflife of $\approx 30 $ hrs. This long lifetime relative to other SHE makes future experiments promising. There is very little experimental or theoretical results for Db with the majority being chemical properties \cite{Schadel2012, Fricke1975}. A calculation of the ionisation potential has been completed for Db in \cite{Dzuba2016} using a relativistic Hartree-Fock approach.

In the $V^{N-1}$ approximation discussed in Section \ref{sec:CIPT} we remove a $7s$ electron.  For the CIPT calculations of Db I we use the same parameters as for the Ta I calculations. The Db I ground state is [Rn]$5f^{14}6d^3 7s^2$ which is similar to Ta I with different principle quantum numbers. For calculation of the even parity states we populated the effective CI matrix with the states of the $6d^3 7s^2$,  $6d^4 7s$ and $6d^5$ configurations. All higher states are obtained through single and double excitations from these states. They are included perturbatively. Similarly for  the states of odd parity the effective matrix contains states of the  $6d^3 7s 7p$, $6d^2 7s^2 7p$ and $6d^4 7p$ configurations. Other configurations are included perturbatively. For the ion  we use the states of the $6d^3 7s$, $6d^2 7s^2$ and $6d^4$ configurations.  Both Breit and radiative corrections are expected to be larger in SHE compared to lighter elements and therefore are included in Table \ref{table:DbISpectrum}. In Table \ref{table:DbISpectrum} we demonstrate the effect of each correction on the energy spectrum of Db I. \\

\begin{table*}[p!]
\center 
\caption{This is the spectrum for the low lying energy levels of Db I using the CIPT method. Here we have included energy levels with and with Briet and QED radiative corrections included. We have also included the ionisation potential of Db I. Of primary importance are the low lying odd states of odd parity which contribute the E1 dipole transitions discussed below. The accuracy of these numbers is expected to be similar to that of Ta  in the period above.  Here $E_{NC}$ are the energies when neither Breit or radiative corrections are included in the calculations, $\Delta_B$ and $\Delta_R$ are the changes in energy from $E_{NC}$ when Breit and radiative corrections are included respectively. The final energy $E$ is the spectrum when both Breit and radiative corrections are included \textit{ab initio}.  \label{table:DbISpectrum}} 
\begin{tabular}{cllcrrrrr}
\toprule
\toprule
  & & & & \multicolumn{4}{c}{Energy} &\\
  \cmidrule{5-8}
& \parbox{2cm}{Major \\ Configuration} & State & $J$ & \parbox{2cm}{No \\ corrections \\ $E_{NC}$ \\ (cm$^{-1}$)} & \parbox{2cm}{Breit \\ correction \\ $\Delta_B$ \\ (cm$^{-1}$)} & \parbox{2cm}{Radiative \\ corrections \\$\Delta_R$\\ (cm$^{-1}$)} &  \parbox{2cm}{Total\\ $E$ \\ (cm$^{-1}$)} & \parbox{1.5cm}{Land\'{e} \\g-factor} \\ 
\midrule
& Even States \\
(1)  &   $6d^3 7s^2$  &  $^4$F  &  $3/2$ & 0 & 0 & 0 & 0 & 0.554 \\ 
(2) &   $6d^3 7s^2$  &  $^4$F &  $5/2$  &  4 072 & -77 & 21 & 4 016 & 1.043 \\ 
(3)  &  $6d^3 7s^2$  &  $^2$F &  $7/2$  &  6 595 & -100 & 31 & 6 527 & 1.170 \\ 
(4)  &  $6d^3 7s^2$  &  $^2$S  & $1/2$ &  7 691 &  -73  &  16 & 7 634 & 2.058 \\ 
(5)  &  $6d^3 7s^2$  &  $^4$G &  $9/2$ &  8 076 & -92 &  33 & 8 017 & 1.191 \\ 
&  Odd States \\
(6) & $6d^2 7s^2 7p$  &  $^2$F$^{\rm_o}$  & $5/2$ &  6 255 & 213 &  123 & 6 591 & 0.739 \\ 
(7)  & $6d^2 7s^2 7p$  &  $^2$D$^{\rm_o}$  & $3/2$ &  11 240 & 156 &  87 & 11 483 &  0.633 \\ 
(8)  &   $6d^2 7s^2 7p$  &  $^2$P$^{\rm_o}$ &  $1/2$ &  12 642 & 140 &  84  & 12 869 & 1.308 \\ 
(9)  &  $6d^2 7s^2 7p$  &  $^4$G$^{\rm_o}$ &  $7/2$  &  13 645 &  116 &  147 & 13 909 & 1.023 \\ 
(10)   &  $6d^2 7s^2 7p$  &  $^4$F$^{\rm_o}$  & $5/2$ &  13 873 & 113  &  132 & 14 117 & 1.067 \\ 
(11)   &  $6d^2 7s^2 7p$  &  $^2$P$^{\rm_o}$  & $1/2$ &  14 516 & 96  &  88  & 14 705 & 0.995 \\ 
(12)   &  $6d^2 7s^2 7p$  &  $^6$F$^{\rm_o}$  & $3/2$ &  14 572 & 105 & 96  & 14 772 & 1.111 \\ 
(13)   &  $6d^2 7s^2 7p$  &  $^4$F$^{\rm_o}$ &  $5/2$ &  17 493 & 78 &76  & 17 647 & 1.111 \\ 
(14)  &  $6d^2 7s^2 7p$  &  $^4$G$^{\rm_o}$  & $9/2$ &  18 596 & 80 &  144  & 18 820  & 1.145 \\ 
(15)  & $6d^3 7s 7p$  &  $^2$D$^{\rm_o}$  & $3/2$ &  19 379 & 62 &  -3  & 19 438 & 0.701 \\ 
 (16)  &  $6d^2 7s^2 7p$  &  $^4$F$^{\rm_o}$ &  $7/2$  &  20 462 & 53 & 134 & 20 649  & 1.203 \\ 
(17)  &  $6d^2 7s^2 7p$  &  $^6$F$^{\rm_o}$  & $3/2$ &  21 706 & 56  & 50 & 21 811 & 1.073 \\ 
(18)  &  $6d^2 7s^2 7p$  &  $^4$D$^{\rm_o}$ &  $1/2$ &  22 123 & 72  & 93 & 22 284 & 0.078 \\ 
(19)   &  $6d^2 7s^2 7p$  &  $^4$F$^{\rm_o}$ &  $5/2$ &  22 204 & 35  & 54 & 22 292 & 1.110 \\ 
(20)  &  $6d^2 7s^2 7p$  &  $^2$D$^{\rm_o}$  & $3/2$ &  23 003 & 39 & 22 & 23 067 & 0.697 \\ 
(21)  &  $6d^2 7s^2 7p$  &  $^2$F$^{\rm_o}$ &  $7/2$  &  23 221 & 37   & 133 & 23 390 & 1.102 \\ 
(22)  &  $6d^3 7s 7p$  &  $^4$F$^{\rm_o}$  & $5/2$ &  23 910 &  4 &  -2 & 23 913 & 0.948 \\ 
(23) &   $6d^2 7s^2 7p$  &  $^2$P$^{\rm_o}$ &  $3/2$ &  24 622 & 2  &  119  & 24 743  & 1.372 \\ 
(24)  &   $6d^2 7s^2 7p$  &  $^2$G$^{\rm_o}$  & $9/2$ &  24 915&  27  & 133 & 25 074 & 1.111 \\ 
 (25)  & $6d^3 7s 7p$  &  $^2$F$^{\rm_o}$ &  $7/2$  &  25 458 & 9  & 17  & 25 480 & 1.152 \\ 
(26) &   $6d^2 7s^2 7p$  &  $^4$F$^{\rm_o}$ &  $5/2$ &  25 510 & 5   & 73 & 25 589 & 1.031 \\ 
(27)   & $6d^2 7s^2 7p$  &  $^2$F$^{\rm_o}$ &  $7/2$ &  26 538 & -4   &  78 & 26 612 & 1.172 \\ 
(28)   &  $6d^2 7s^2 7p$  &  $^2$S$^{\rm_o}$ &  $1/2$ &  27 435 & -10   &  49 & 27 479 & 1.663 \\ 
(29)  &   $6d^2 7s^2 7p$  &  $^2$F$^{\rm_o}$ &  $7/2$  &  27 662 & -23   & 24  & 27 666 & 1.128  \\ 
(30)  &  $6d^2 7s^2 7p$  &  $^4$D$^{\rm_o}$ &  $3/2$ &  27 589 & -5   & 114 & 27 697  & 1.147 \\ 
(31)  &  $6d^2 7s^2 7p$  &  $^4$G$^{\rm_o}$ &  $9/2$ &  27 885 & -13   &  118 & 27 990 & 1.173 \\ 
(32)   & $6d^2 7s^2 7p$  &  $^2$D$^{\rm_o}$ &  $5/2$ &  28 162 & -25  & 75 & 28 211  & 1.130 \\ 
(33)    &  $6d^2 7s^2 7p$  &  $^4$P$^{\rm_o}$ &  $3/2$ &  29 183 & 1  & 74 & 29 259 & 1.659 \\ 
(34)  & $6d^2 7s^2 7p$  &  $^4$G$^{\rm_o}$ &  $11/2$ &  29 669 & -45 & 103 & 29 669 & 1.254 \\ 
(35)  & $6d^3 7s 7p$  &  $^6$G$^{\rm_o}$ &  $9/2$ &  29 946 & -75  & -87 & 29 784 & 1.254 \\ 
(36)  &   $6d^2 7s^2 7p$  &  $^6$F$^{\rm_o}$  & $5/2$ &  29 734 &  -25  & 174  & 29 821 & 1.343 \\ 
(37) &  $6d^3 7s 7p$  &  $^4$D$^{\rm_o}$ &  $1/2$ &  29 886 & -24  &  -33  & 29 832 & 0.220 \\ 
(38)  &  $6d^2 7s^2 7p$  &  $^2$F$^{\rm_o}$  &  $7/2$ &  30 474 & -29  & 97 & 30 541  & 1.161 \\ 
 & Db II states\\
(39)  &   $6d^2 7s^2$  &  $^3$F &  $2$ & 56 546 & 48 & 139 & 56 733 & 0.731 \\ 
(40)  &   $6d^2 7s^2$  &  $^3$S &  $0$ & 62 673 & -13 & 119 & 62 778 & 0.000 \\ 
(41)  &   $6d^2 7s^2$  &  $^3$F &  $3$ & 62 952 & -45 & 176 & 63 083 & 1.083 \\ 
(42)  &   $6d^2 7s^2$  &  $^3$D &  $2$ & 65 122 & -62 & 120 & 65 179 & 1.250 \\ 
(43)  &   $6d^2 7s^2$  &  $^3$P &  $1$ & 65 587 & -79 & 113 & 65 620 & 1.467 \\ 
(44)  &   $6d^2 7s^2$  &  $^5$G &  $4$ & 67 466 & -83 & 189 & 67 572 & 1.120 \\ 
\bottomrule 
 \bottomrule 
\end{tabular} 
\end{table*}
Comparing the Db I spectrum in Table \ref{table:DbISpectrum} to the Ta I spectrum in Table \ref{tab:TaComparison} we see that the order of the even parity states has remained the same relative to each other. However the order of the odd states has been significantly altered with the first $2F^{o}_{5/2}$ state being significantly lowered in the spectrum. Another thing to note is that the odd parity excitations are typically $6d \rightarrow 7p$ as opposed to the Ta excitation $6s \rightarrow 6p$. This can be explained by relativistic effects where the $7s$ electrons are more tightly bound than the $6d$ electrons in contrast to the $5d$ and $6s$ electrons in Ta. These relativistic effects also cause the $6d$ electron to be ionised in Db instead of the $6s$ electron. This may result in significantly different chemical properties in Db compared to Ta. \\

 From Table \ref{table:DbISpectrum} we see that the effect of both the Breit interaction ($\Delta_B$) and radiative corrections ($\Delta_R$)  is small and lies within the accuracy of our code with the maximum correction being less than 200 cm$^{-1}$. Interestingly we see, to the accuracy of our calculations, that the effects are linear and do not correlate with each other. This can be seen by summing the two  corrections and the calculated energy with no  corrections included ($E_{NC}$). This energy is very close to states in the spectrum which includes both corrections simultaneously,
\begin{align*}
E \approx E_{NC} + \Delta_B + \Delta_R.
\end{align*}
To test the consistency of our method we calculated the spectrum of Db I using the $V^{N-1}$ approach removing a $6d$ electron for the frozen core potential. In these calculations we obtained a similar spectrum within the accuracy of our calculations.

We are not aware of any other calculations of the Db spectrum apart from the calculations of IP. Out value 56744~cm$^{-1}$ is in good agreement with the Hartree-Fock number 55000(7000)~cm$^{-1}$~\cite{Dzuba2016} and
coupled cluster number 55590~cm$^{-1}$~\cite{BorschevskyPC}.

\section{Electric dipole transitions and isotope shift} \label{sec:E1transitions}

Of particular experimental interest in calculating the spectra of heavy elements are the electric dipole (E1) transitions. In this work we calculate and present the E1 transition amplitudes for the major optical  transitions from the ground state to the  lowest lying odd parity states for each Ta I and Db I. It should be noted that there is no published data for the E1 transitions for either Ta I or Db I and therefore we present the E1 transition amplitudes ($A_{E1}$)  and transition probabilities ($T_{E1}$) for both atoms. \\

To calculate the E1 transition amplitudes $A_{E1}$ we use the self-consistent random-phase approximation (RPA) to simulate the atom in an external electromagnetic field. This results in an effective electric dipole field for the electrons. The E1 transition amplitude for a transition between states $a$ and $b$ is given by $A_{E1} = \left< b || \hat D + \delta V || a\right>$ where $|a>$ and $|b>$ are the many electron wavefunctions calculated in the CIPT method above, $\hat D$ is the electric dipole operator acting on external electrons, $\delta V$ is the correction to the self-consistent Hartree-Fock potential of atomic core caused by photon electric field. For a more in depth discussion on this method refer to ref. \cite{Dzuba2018}. \\

The E1 transition rates are calculated using (in atomic units),
\begin{align*}
T_{E1} = \dfrac{4}{3}\left(\alpha \omega\right)^3\dfrac{ A_{E1}^2}{2J + 1}
\end{align*}
where $J$ is the angular momentum of the upper state, $\alpha$ is the fine structure constant and $\omega$ is the frequency of the transitions in atomic units. All calculations obey the selection rules for E1 transitions, a change in parity and change in angular momenta $|\Delta J| \leq 1 $. We present the E1 transitions for Ta I and Db I in Table \ref{table:E1Amplitudes}.\\

\begin{table*}[t!] 
\caption{This table presents the single electric dipole transisitions from the  ground state of Db I ($^4Fo_{3/2}$) and Ta I ($^4F_{3/2}$) to the low lying odd parity states. These low lying optical transition obey the E1 transition selection rules with a change of parity and change of angular momentum $|\Delta J| < 1$ where the leading contribution is for the odd parity state. The numbers next to the states correspond to the numbered spectra in Tables \ref{tab:TaComparison} and \ref{table:DbISpectrum}. The transition amplitudes $A_{E1}$ are in atomic units. For the Db I transitions we include the associated isotope shift parameters $a$ and $F$. The isotope shift calculation was performed for $^{268}$Db ($\left<r^{2}\right>_{268} = 36.770$ fm$^2$) and $^{289}$Db ($\left<r^{2}\right>_{289}=38.470$ fm$^2$).\label{table:E1Amplitudes} }
\begin{tabular}{cl@{\hspace{0.75cm}}r@{\hspace{0.75cm}}r@{\hspace{0.75cm}}|ccr@{\hspace{0.5cm}}r@{\hspace{0.5cm}}r@{\hspace{0.5cm}}r}  
\toprule
\toprule
\multicolumn{4}{c}{Ta I} & \multicolumn{6}{c}{Db I} \\ \\
 & State &   \parbox{1cm}{$A_{E1}$ \\ (a.u)} & \parbox{1.5cm}{$T_{E1}$ \\ $(\times 10^{6} \ \text{s}^{-1})$  } & & State &   \parbox{2cm}{$A_{E1}$ \\ (a.u)} & \parbox{1.5cm}{$T_{E1}$ \\ $(\times 10^{6} \ \text{s}^{-1})$  } &  \parbox{1cm}{$a  $ (cm$^{-1}$)} & \parbox{1.5cm}{$F $ \\ (cm$^{-1}$/fm$^{2}$)} \\
\midrule
 (6) & $^6$G$^{\rm_o}_{3/2}$  & -0.270   & 0.194 & (6) & $^2$F$^{\rm_o}_{5/2}$  &   0.631  & 0.0385 & 32.16 & 3.11  \\
(7) & $^2$F$^{\rm_o}_{5/2}$  & 0.214 &  0.090 & (7)  & $^2$D$^{\rm_o}_{3/2}$ & 1.53  &  1.80 & 18.70 & 1.81  \\
(8) & $^4$D$^{\rm_o}_{1/2}$   & -0.641   & 2.64 & (8) &  $^2$P$^{\rm_o}_{1/2}$  & 0.558 &  0.672 & -3.42 & -0.33 \\
 (9) & $^6$G$^{\rm_o}_{5/2}$  & -0.434   & 0.449 & (10) & $^4$F$^{\rm_o}_{5/2}$    & -0.531 &  0.268 & 27.33 & 2.64  \\
 (10) & $^4$D$^{\rm_o}_{3/2}$   & 0.149   &  0.0856 & (11) & $^2$P$^{\rm_o}_{1/2}$    & 0.384 & 0.476 & 15.78 & 1.52  \\
 (11) &$^2$S$^{\rm_o}_{1/2}$    & -0.107  & 0.0973 &(12)  & $^6$F$^{\rm_o}_{3/2}$ & 0.180 &  0.0527 & 14.93 & 1.44  \\
 (13) &$^2$D$^{\rm_o}_{3/2}$    & 0.495   & 1.12 & (13) & $^4$F$^{\rm_o}_{5/2}$  &    -0.339 & 0.213 & 8.39 & 0.81 \\
 (14) &$^4$D$^{\rm_o}_{5/2}$   & -0.200   &  0.128 & (15) & $^2$D$^{\rm_o}_{3/2}$  &   -0.343  & 0.437 & -18.84 & -1.82    \\
 (15) &$^4$F$^{\rm_o}_{3/2}$  & -0.360   & 0.688 & (17)  & $^6$F$^{\rm_o}_{3/2}$  &    1.22  & 7.85 & -0.33 & -0.03  \\
 (16) &$^2$D$^{\rm_o}_{5/2}$    & 0.069   & 0.0160 & (18) & $^4$D$^{\rm_o}_{1/2}$  &    0.0968 & 0.105 & 13.58 & 1.31  \\
 (19) &$^6$F$^{\rm_o}_{1/2}$   &   0.019   & 0.00446 & (19) & $^4$F$^{\rm_o}_{5/2}$ &    -0.163 & 0.0996& -1.54 & -0.51 \\
 (20) &$^4$F$^{\rm_o}_{5/2}$   &  -0.094   & 0.0381 & (20) & $^2$D$^{\rm_o}_{3/2}$  &    0.784 & 3.83 & -4.88 & -0.47 \\
 (22) &$^6$F$^{\rm_o}_{3/2}$  & 0.007   & 0.000412 & (22) & $^4$F$^{\rm_o}_{5/2}$  &    -1.01 & 4.70  & -19.24 & -1.86 \\
 (23) &$^6$D$^{\rm_o}_{1/2}$    & -0.073   & 0.0795 & (23)  & $^2$P$^{\rm_o}_{3/2}$  &   -0.150 & 0.173 & 16.75 & 1.62 \\
 (24) &$^6$D$^{\rm_o}_{3/2}$ & -0.249   & 0.477 & (26) & $^4$F$^{\rm_o}_{5/2}$  &    -0.890 & 4.49  & 6.22 & 0.60 \\
 (27) &$^6$F$^{\rm_o}_{5/2}$   & -0.356   & 0.683 & (28) & $^2$S$^{\rm_o}_{1/2}$ &     -0.570 & 6.83 & -4.42 & -0.43 \\
 (29) &$^4$D$^{\rm_o}_{1/2}$   &  0.282   & 1.34 & (30) & $^4$D$^{\rm_o}_{3/2}$ &   -0.114 & 0.139 & 16.04 & 1.55  \\
 (31) &$^4$P$^{\rm_o}_{5/2}$ & 0.202   & 0.248 &  (32) & $^2$D$^{\rm_o}_{5/2}$ &  0.228 & 0.393 & 3.31 & 0.32 \\
 (32) &$^4$D$^{\rm_o}_{3/2}$   & 0.405   & 1.53 & (33)  & $^4$P$^{\rm_o}_{3/2}$ &   -0.388 & 2.01 & 6.68 &  0.64  \\
 (34) &$^4$P$^{\rm_o}_{3/2}$   &  -0.063   & 0.0377&  (36) & $^6$F$^{\rm_o}_{5/2}$ &    -0.0174   & 0.00270 & 14.86 & 1.44 \\
 (35) &$^6$D$^{\rm_o}_{5/2}$  & 0.338   & 0.741 & (37)  & $^4$D$^{\rm_o}_{1/2}$  &   1.49 & 59.7 & -28.14 & -2.72  \\
 (36) &$^4$P$^{\rm_o}_{1/2}$    & -0.066   & 0.0859 &   \\
 (41) &$^6$D$^{\rm_o}_{5/2}$  & -0.278   & 0.583 & \\
 (43) & $^2$P$^{\rm_o}_{1/2}$  &  -0.295   & 1.04 & \\
\bottomrule
\bottomrule
\end{tabular}
\end{table*}

For Db from Table \ref{table:E1Amplitudes} we see that the  transitions from the ground state with the largest transition rates are to the odd parity state $^4F_{3/2} \rightarrow ^4D^{o}_{1/2}$.  A reason for these large E1 amplitudes for these states could be due to the larger contribution of the $7p \rightarrow 7s$ transition as opposed to the more suppressed $7p \rightarrow 6d$ transition.

Finally, we calculate isotope shift for Db. Isotope shift is important since it helps to obtain information about
nuclei of SHE when frequencies of the transitions are measured for several isotopes. It can also be used 
to predict the spectra of other isotopes, in particular the spectrum of the hypothetically stable neutron-rich 
isotopes with ``magic'' number of neutrons $N=184$. This may help in search for such isotopes.

Isotope shift of SHE elements is strongly dominated by volume shift (also known as ``field shift'' in literature). We calculate it by varying nuclear radius in 
computer codes. We present results in two different forms. First is given by~\cite{DFW17}
\begin{align*}
\delta \nu &= E_{2} - E_{1} = a\left(A_{2}^{1/3} - A_{1}^{1/3}\right),
\end{align*}
where $A_1$ and $A_2$ are atomic numbers for two isotopes ($A_2>A_1$) and $a$ is the parameter which
comes from the calculations. This form is convenient for prediction of the spectra of heavier isotopes. It is motivated by the relativistic dependence of the volume shift on the nuclear radius, $R_N$, which is proportional to $R_N^{2\gamma}$ where $\gamma = \sqrt{1 - (Z\alpha)^2}$. For  Db  $R_N^{ 2\gamma}  \approx R_{N}^{1.28}$ and using the large scale trend for nuclear radii $R_N \propto A^{1/3}$  the volume shift can be approximated by $\propto A^{1/3}$. This nuclear radius approximation is valid for large scale trends in $A$ where nuclear shell fluctuations are suppressed \cite{Angeli2013, DFW17}, this is applicable for our Db I calculations as $A_1$ and $A_2$ are not neighboring isotopes.

Another form for the isotope shift is the standard formula related the change of atomic frequency to the change
of nuclear radius
\begin{align*}
\delta \nu &= F\delta \left<r^{2}\right>.
\end{align*}
This formula is convenient for extraction of the nuclear radius change from the isotope shift measurements.
The values of the $a$ and $F$ parameters for strong electric dipole transitions of Db are presented in 
Table~\ref{table:E1Amplitudes}.

\section{Conclusion}

We have calculated energy levels, electric dipole transition amplitudes and isotope shift for superheavy element
dubnium. Similar calculations for its lighter analog Ta indicate that the uncertainty of the results for the energies of 
Db is unlikely to exceed 500~cm$^{-1}$. Db is the first SHE with open $6d$ shell which is studied with the recently
developed CIPT method. The successful use of the CIPT method for Db opens a way to perform similar study for 
all SHE with open $6d$ shell up to Mt ($Z=109$). 

\acknowledgments

We thank Julian Berengut and Daniel Czapski for useful discussions.
This work was funded in part by the Australian Research Council.

\bibliographystyle{apsrev4-1}
\bibliography{SHE}

\end{document}